\documentclass[11pt]{article}

\usepackage{fullpage}

\usepackage{latexsym}
\usepackage{amssymb}
\usepackage{amsmath}
\usepackage{enumerate}
\usepackage{mathtools}
\usepackage{verbatim}
\usepackage{color}

\usepackage{hyperref}

\def\e{\mbox{$\epsilon$}}

\def\X{\mbox{$\cal X$}}

\def\B{\mbox{$\cal B$}}
\def\N{\mbox{$\mathbb N$}}

\def\L{\mbox{$\cal L$}}

\def\E{\mbox{$\cal E$}}

\def\F{\mathbb F}

\def\R{\mathbb R}

\def\cmp{\mbox{$\mathbb C$}}

\def\e{\mbox{$\epsilon$}}

\def\E{\mbox{$\cal E$}} 
 
\def\F{\mathbb F}
\def\Z{\mathbb Z}

\def\Z{\mathbb Z}

\def\finito{{\hspace*{\fill}  \mbox{$\blacksquare $}}}

\def\defeq{:=} 

\def\ceil#1{\left\lceil #1 \right\rceil}
\def\floor#1{\left\lfloor #1 \right\rfloor}

\newtheorem{theorem}{Theorem}[section]
\newtheorem{lemma}[theorem]{Lemma}
\newtheorem{definition}[theorem]{Definition}
\newtheorem{corollary}[theorem]{Corollary}

\def\X{\chi}

\def\e{{{\epsilon}}}
\def\al{{{\alpha}}}

\def\EX{\mathbb E} 

\def\B{\mbox{$\cal H$}}

\begin{document}
\author{Louay Bazzi
\footnote{
  Department of Electrical and Computer Engineering, American University of Beirut, 
Beirut, Lebanon. E-mail: louay.bazzi@aub.edu.lb.} 
}
\title{On the tightness of Tiet\"{a}v\"{a}inen's bound for distributions with limited independence} 

\maketitle
\abstract{ 
In 1990, Tiet\"{a}v\"{a}inen showed that if the only information
we know about a linear code is its dual distance $d$, then its covering radius $R$ is at most  $\frac{n}{2}-(\frac{1}{2}-o(1))\sqrt{dn}$. 
  While Tiet\"{a}v\"{a}inen's bound was later improved for large values of $d$, 
  it is still the best known upper bound for  small values including the $d = o(n)$ regime.
  Tiet\"{a}v\"{a}inen's bound holds also for $(d-1)$-wise independent  probability  distributions on $\{0,1\}^n$,  of which 
linear codes with dual distance $d$ are special cases.   
We show that    Tiet\"{a}v\"{a}inen's bound on $R-\frac{n}{2}$ is asymptotically  
tight up to a factor of $2$ for $k$-wise independent distributions 
if $k\leq\frac{n^{1/3}}{\log^2{n}}$. Namely, we show that 
there exists  a $k$-wise independent probability distribution $\mu$  on $\{0,1\}^n$
whose covering radius is at least  $\frac{n}{2}-\sqrt{kn}$.    
Our key technical contribution is the following     lemma on low degree polynomials, which implies 
the existence of $\mu$ by linear programming duality.
We show that,   for sufficiently large 
  $k\leq\frac{n^{1/3}}{\log^2{n}}$ and   for each  polynomial $f(v)\in \R[v]$ of degree at most $k$, the
  expected value of $f$ with respect to the binomial distribution cannot be positive  if 
   $f(w)\leq 0$  for each integer $w$  such that 
  $|w-n/2|\leq\sqrt{kn}$. The proof  uses tools from approximation theory.          
}
    

\section{Introduction}\label{intros}

The {\em covering radius} of  a subset $C$ of the Hamming cube $\{0,1\}^n$  is the minimum $R$  such that any vector in $\{0,1\}^n$ is within Hamming distance at most $R$
from $C$. 
Studying the relation between the covering radius of a binary linear  code and its   dual code goes back to  Delsarte \cite{Del73a}  
(see also Helleseth, Kl{\o}ve, and  Mykkeltveit \cite{HKM78} and  Sole \cite{Sole90}).  For a general reference on covering codes, see  
Cohen, Honkala, Litsyn, and  Lobstein's book \cite{CC97}. 

Based on Delsarte  linear programming relaxation \cite{Del73}, 
Tiet\"{a}v\"{a}inen showed in 1990    that if the only information
we know about a linear code $C $ is its dual distance $d$, then its covering radius $R$ cannot be too large: 
\begin{theorem}[Tiet\"{a}v\"{a}inen    \cite{T90,T91}]\label{trth90}
      {\bf (Upper bound on the covering radius of codes in terms of  dual distance)} 
  Let  $C\subset \F_2^n$ an $\F_2$-linear code whose dual has minimum distance  $d\geq 2$.
  Then the covering   radius $R$ of $C$ is at most
  \[
  \left\{\begin{array}{ll}
    \frac{n}{2} - \sqrt{s(n-s)} + s^{1/6} \sqrt{n-s} & \mbox{if $d=2s$ is even} \\
    \frac{n}{2} - \sqrt{s(n-1-s)} + s^{1/6} \sqrt{n-1-s} - \frac{1}{2} & \mbox{if $d=2s+1$ is odd.} 
  \end{array}\right.
  \]
     \end{theorem}
Tiet\"{a}v\"{a}inen's bound was later improved in the $d = \Theta(n)$ regime  in a sequence of works 
\cite{SS93} - \cite{AB02} by  Sole, Stokes, Honkala, Litsyn,  Tiet\"{a}v\"{a}inen,    Struik,   Honkala,   Laihonen,   Ashikhmin, and    Barg. 
See also  
Fazekas and Levenshtein \cite{FL95} for extensions to polynomial metric spaces and
Chapters 8 and 12 in  \cite{CC97}. 

For sufficiently small values of $d =\Theta(n)$,  Tiet\"{a}v\"{a}inen's bound is still the best known upper bound on the covering radius as a function of dual distance. Actually, Tiet\"{a}v\"{a}inen argued  in \cite{T91} that improving his bound in the $d=o(n)$ regime is difficult since this regime  includes dual BCH codes and accordingly 
improvements  would give new interesting results on character sums.

The focus of this paper is on $d = o(n)$, i.e., on rate-zero linear codes of subexponential size.
A natural question is how  
tight Tiet\"{a}v\"{a}inen's bound is in this regime. 
That is, if $d$ is sub-linear in $n$, what can we say about the covering radius of a code given only its dual distance $d$?

  As noted by Tiet\"{a}v\"{a}inen   \cite{T90}, we  know from dual BCH codes that  
  if $n = 2^m -1$,  where  $m\geq  2$ and $s\geq 1$ are integers such that $s < \frac{1}{2}\sqrt{n+1}+1$, then there are
  codes with dual distance $2s+1$ and covering radius $R$ satisfying the lower bound
    \footnote{ 
    The lower bound in (\ref{bchlb}) follows immediately from 
    Weil-Carlitz-Uchiyama's bound (it is  also 
    slightly better than the lower bound $R\geq \frac{n}{2} - s\sqrt{n+1}$ 
    stated on p. 1473 in \cite{T90}). Let  $n = 2^m -1$, where  $m\geq 2$ an integer,  and
    let $s\geq 1$ be an integer such that $2s-2< 2^{m/2}$, i.e., $s < \frac{1}{2}\sqrt{n+1}+1$. 
    Weil-Carlitz-Uchiyama's bound (see \cite{MS77}) asserts that        for each non-zero codeword $x\in BCH(s,m)^\bot$, 
      we have $||x|-2^{m-1}|\leq (s-1)2^{m/2}$. 
      Thus,  (\ref{bchlb}) holds because  the all-ones vector  $\vec{1} \not\in BCH(s,m)^\bot$ because $n$ is odd and
    $\vec{1}\in BCH(s,m)$.}
\begin{equation}\label{bchlb}
  R\geq \frac{n}{2} - (s-1)\sqrt{n+1}-\frac{1}{2}. 
  \end{equation}
Asymptotically, the lower bound on $R - \frac{n}{2}$ in (\ref{bchlb}) is away from Tiet\"{a}v\"{a}inen's bound by a $\sqrt{\frac{d}{2}}$ factor, which is
considerable  for $d = w(1)$.

  While we  do not resolve in this paper the question of tightness of 
  Tiet\"{a}v\"{a}inen's bound for linear codes  in the $d = o(n)$ regime,   we show that it is essentially tight for the bigger class of $k$-wise independent distributions in the $k \leq \frac{n^{1/3}}{\log^2{n}}$ regime.

  A probability distribution $\mu$ on $\{0,1\}^n$ is called {\em $k$-wise independent} if sampling $x\sim \mu$ gives a random vector $x=(x_1,\ldots, x_n)$,
where  each $x_i$ is equally likely to be $0$ or $1$ and any $k$ of the $x_i$'s are statistically independent   \cite{Lub85,Vaz86}.
Linear codes with dual distance $d$ are special cases of  $k$-wise independent  probability distributions on $\{0,1\}^n$, where 
$k=d-1$;  if $\mu$ is a probability distribution on $\{0,1\}^n$ uniformly distributed on an $\F_2$-linear code $C\subset \F_2^n$,  
  then $\mu$ being $k$-wise independent is equivalent to $C$ having
  dual minimum distance at least $k+1$.   
  If $\mu$ is a probability distribution on $\{0,1\}^n$, define the {\em covering radius} of  $\mu$ to be  the covering radius of its support.
  
Tiet\"{a}v\"{a}inen's bound is based on the following lemma which asserts the existence of certain low degree polynomials. 
     Let $B_n$ be the binomial distribution on $[0:n]:= \{0, \ldots, n\}$,  i.e., $B_n(w): =\frac{1}{2^n}{\binom{n}{w}}$.    
     \begin{lemma}[Tiet\"{a}v\"{a}inen         \cite{T90,T91}]\label{trl90} {\bf (Low degree polynomials lower bound)}
  Let $1 \leq k \leq n-1$ be integers. There exits a polynomial $p(v)\in \R[v]$ of degree at most $k$ such that $\EX_{B_n} p>0$ and  $p(w)\leq 0$, for each $w\in [0:n]$ such that
  \[
  \left\{\begin{array}{ll}
    w\geq \floor{\frac{n}{2} - \sqrt{s(n-s)} + s^{1/6} \sqrt{n-s}}+1 & \mbox{if $k=2s-1$ is odd} \\
    w\geq \floor{\frac{n}{2} - \sqrt{s(n-1-s)} + s^{1/6} \sqrt{n-1-s} - \frac{1}{2}}+1 & \mbox{if $k=2s$ is even.} 
  \end{array}\right.  
  \]
     \end{lemma}
     Tiet\"{a}v\"{a}inen established his bound using Krawtchouk polynomials. 
It is not hard to see that Lemma \ref{trl90} actually shows more than Theorem \ref{trth90}; it gives the following upper bound on the covering radius of $k$-wise independent distributions:
  \begin{corollary}\label{trc90}
    {\bf (Upper bound on the covering radius of $k$-wise independent distributions)} 
Let $1\leq k \leq n-1$ be integers and let $\mu$ be  a  $k$-wise independent probability distribution on $\{0,1\}^n$. 
  Then the covering  radius of $\mu$ is at most
  \[
  \left\{\begin{array}{ll}
    \frac{n}{2} - \sqrt{s(n-s)} + s^{1/6} \sqrt{n-s} & \mbox{if $k=2s-1$ is odd} \\
    \frac{n}{2} - \sqrt{s(n-1-s)} + s^{1/6} \sqrt{n-1-s} - \frac{1}{2} & \mbox{if $k=2s$ is even.} 
  \end{array}\right.
  \]
\end{corollary} 
  Actually, Corollary \ref{trc90} is equivalent to Lemma \ref{trl90}.
First, we note that this follows from the linear programming  duality between low degree polynomials and $k$-wise independent distributions: 
      \begin{lemma}\label{covlpduality}{\bf (Duality between  low degree polynomials and $k$-wise independence distributions)} 
    Let $1\leq k\leq n$ be integers and $R>0$ a real number. Then   the following are equivalent:
  \begin{itemize}
  \item[I)]  Each $k$-wise independent probability distribution on $\{0,1\}^n$ has  covering radius
    less than $R$
   \item[II)] There exits a polynomial $p(v)\in \R[v]$ of degree at most $k$ such that $\EX_{B_n} p>0$ and 
          $p(w)\leq 0$, for each $w\in [0:n]$ such that $w\geq R$.  
   \end{itemize}
  \end{lemma}
  The implication from (II) to (I) was implicitly used by
  Tiet\"{a}v\"{a}inen   in his  proof of Theorem \ref{trth90} in the context of linear codes. 

   We show that,     for $k$-wise independent distributions,
Tiet\"{a}v\"{a}inen's bound on $R -\frac{n}{2}$ 
   is asymptotically tight up to  a factor of $2$ if $k \leq \frac{n^{1/3}}{\log^2{n}}$: 
      \begin{theorem} \label{mexi}
{\bf (Lower bound on the covering radius of $k$-wise independent distributions)} 
        There exist   absolute constants $k_0,n_0>0$ such that for each integer $n\geq n_0$ and each integer $k$ satisfying $k_0 \leq k \leq \frac{n^{1/3}}{\log^2{n}}$, there exists a $k$-wise independent probability distribution on $\{0,1\}^n$ whose covering radius is at least  $\frac{n}{2}-\sqrt{kn}$.         
      \end{theorem}
      The key technical contribution of the this paper  
      is the following result about low degree polynomials. 
\begin{theorem}\label{mainth} {\bf (Low degree polynomials upper bound)}
  There exist   absolute constants $k_0,n_0>0$ such that for each integer $n\geq n_0$ and each integer $k$ satisfying $k_0 \leq k \leq \frac{n^{1/3}}{\log^2{n}}$, the following holds. 
For each  polynomial $f(v)\in \R[v]$ satisfying 
\begin{itemize}
\item[i)] $deg(f)\leq k$
\item[ii)] $f(w)\leq 0$, for each integer $w\in [0:n]$ such that 
  $|w - n/2|\leq \sqrt{kn}$,
\end{itemize}
we must have $\EX_{B_n} f\leq 0$.
\end{theorem}
By further constraining  (ii) in Theorem \ref{mainth}, we get the following. 
\begin{corollary}\label{mainthc} {\bf (Low degree polynomials upper bound)}
There exist   absolute constants $k_0,n_0>0$ such that for each integer $n\geq n_0$ and each integer $k$ satisfying $k_0 \leq k \leq \frac{n^{1/3}}{\log^2{n}}$, the following holds.  
For each  polynomial $f(v)\in \R[v]$ satisfying 
\begin{itemize}
\item[i)] $deg(f)\leq k$
\item[ii)] $f(w)\leq 0$, for each integer $w\in [0:n]$ such that $w\geq n/2-\sqrt{kn}$,
\end{itemize}
we must have $\EX_{B_n} f\leq 0$.
\end{corollary}
Thus,  Theorem \ref{mexi} follows from Corollary \ref{mainthc} via the  duality in Lemma \ref{covlpduality}.

The proof of Theorem \ref{mainth}     uses tools from approximation theory. 
At high level, we will bound $\EX_{B_n} f$ by examining the values of $f$ on integer sequences of 
length $k+1$  contained in the interval of points $w\in [0:n]$ such that $|w - n/2|\leq \sqrt{kn}$.
The sequence are   disjoints  and they have small 
Lebesgue constant. We will show that each sequence contains a  point on which the negative value of $f$ is large in absolute value (assuming that $f$ is not identically zero). Those points will be used to show that $\EX_{B_n} f \leq 0$.  The sequences will be constructed from translates of a quantized Chebyshev sequence whose Lebesgue constant will be estimated using Markov's theorem.   

The use of approximation theory tools in the proof  was  inspired by the works of  Paturi \cite{Pat92} and Linial and Nissan \cite{LN90}.
Paturi implicitly used  the Lebesgue constant of equally-spaced sequences and he  used Markov's theorem to  estimate the approximate degree of symmetric boolean functions. 
Linial and Nissan   used properties of quantized zeros of Chebyshev polynomials to approximate the inclusion-exclusion formula. 
At a high level, the new ingredient in our argument is the use of multiple sequences and in particular the translated  sequences technique.

Another related work which builds on \cite{Pat92}  is the author's joint work with Nahas  on read-once CNF formalas and small-bias spaces \cite{Baz15}.  See also the 
aforementioned papers  \cite{HKT95}, \cite{LT96},
\cite{HLL97} - \cite{LS99} which 
use  Chebyshev polynomials to improve on Tiet\"{a}v\"{a}inen's bound in the $d = \Theta(n)$ regime.

\section{Paper outline}

After summarizing the  notations and terminology used throughout the paper in  Section \ref{termo}, we prove Lemma \ref{covlpduality} 
in  Section  \ref{covlpdualitypms}.  
The proof of Theorem \ref{mainth} uses tools from approximation theory, which we  explain in Sections \ref{appmach} and \ref{std}.
After  explaining the proof technique and outline in Section \ref{prtech}, we establish  Theorem \ref{mainth} in 
Sections \ref{chebybound} and  \ref{concprf}.  
We conclude in Section \ref{conc} with  open questions. 

\section{Preliminaries}\label{termo}

The following section summarizes the terminology used in the paper. Section \ref{frprelS} contains Fourier analysis notions on the hypercube used in the proof
of Lemma \ref{covlpduality}.

\subsection{Terminology}
Throughout the paper, $n\geq 1$ is an integer. We will use the following notations. 
If $x \in \{0,1\}^n$, the {\em Hamming weight} of $x$, which we  denote by $|x|$,  is the  number of nonzero  coordinates of $x$.
The set $\{0, \ldots, n\}$ is denoted by $[0:n]$.  
The {\em binomial distribution}  on $[0:n]$ is denoted by $B_n$, i.e., $B_n(w)=\frac{1}{2^n}{\binom{n}{w}}$.
The {\em uniform distribution}  on $\{0,1\}^n$ is denoted by $U_n$, i.e., $U_n(x) = \frac{1}{2^n}$ , for all $x\in \{0,1\}^n$. 
The finite field  structure on $\{0,1\}$  is denoted by $\F_2$.   
The {\em minimum distance} of a non-empty $\F_2$-linear code 
is the minimum weight of a nonzero codeword.  Throughout this paper, $\log$ means $\log_e$.

If $\mu$ is a  probability distribution,  $\EX_\mu$  denotes the expectation  with respect to $\mu$    and ``$x\sim \mu$''  denotes the process of sampling  a random point $x$ according to $\mu$.   
A probability distribution $\mu$ on $\{0,1\}^n$ is called {\em $k$-wise independent} if sampling $x\sim \mu$ gives a random vector $x=(x_1,\ldots, x_n)$,
where  each $x_i$ is equally likely to be $0$ or $1$ and any $k$ of the $x_i$'s are statistically independent   \cite{Lub85,Vaz86}.  See also Section \ref {frprelS} for an equivalent definition. 
 
If $r\geq 0$ is a real number and $x\in \{0,1\}^n$,  $\B_n(x; r)$ denotes the  radius-$r$ {\em Hamming ball} in $\{0,1\}^n$  
centered at $x$, i.e.,
$\B_n(x; d) = \{x\in \{0,1\}^n~:~ |x+y|\leq r\}$. If $C$ is a subset of $\{0,1\}^n$, $\B_n(C; r)$ denotes the
$r$-neighborhood of $C$ with the respect to the Hamming distance, i.e., $\B_n(C; r) = \cup_{x\in C} \B_n(x; d)$.  
  The {\em covering radius} of  $C$ 
  is the minimum $r$
  such that $\B_n(C; r) = \{0,1\}^n$.
  Equivalently,    the covering radius of  $C$ is the minimum $r$ 
  such that $\B_n(x; r) \cap C \neq \emptyset$ for each $x\in \{0,1\}^n$. 
  If $\mu$ is a probability distribution on $\{0,1\}^n$, the {\em covering radius} of  $\mu$ is the covering radius of its support.
  Equivalently,  the covering radius of $\mu$ is the minimum $r$
  such that $\mu(\B_n(x; r))\neq 0$ for each $x\in \{0,1\}^n$.

\subsection{Fourier transform preliminaries}\label{frprelS}
  The use of   Harmonic analysis  methods in coding theory dates back to   MacWilliams \cite{Mac63}. 
  We give below  some preliminary notions used in the proof of Lemma \ref{covlpduality}; see also  \cite{Baz15}.

Identify the hypercube $\{0,1\}^n$ with the abelian 
group  $\Z_2^n = (\Z/2\Z)^n$ and  consider the   {\em characters} $\{ \X_z \}_{z\in \Z_2^n}$ 
of $\Z_2^n$, 
 where  $\X_z:\lbrace 0,1 \rbrace^n \rightarrow \{-1,1\}$ is given by $\X_z(x) =  (-1)^{\langle x, z\rangle}$  and 
$\langle x, z\rangle = \sum_{i=1}^{n} x_iz_i$. 
Consider the 
$\cmp$-vector space $\L(\Z_2^n)$  of complex valued functions defined on $\Z_2^n$  and consider the 
 inner product on $\L(\Z_2^n)$:   $$\langle f,g\rangle = \EX_{U_n} f \overline{g}= \frac{1}{2^n} \sum_{x} f(x) \overline{g(x)}.$$ 
 The characters $\{ \X_z \}_z$ form an orthonormal  basis of $\L(\Z_2^n)$,
 i.e., $\langle \X_z,\X_{z'}\rangle = \delta_{z,z'}$,   for each $z,z'\in \{0,1\}^n$,  where $\delta$ is the Kronecker  delta function.
 
If $f\in \L(\Z_2^n)$, its Fourier transform 
$\widehat{f}\in \L(\Z_2^n)$ is given by the coefficients of the unique 
 expansion of $f$ in terms of the characters: \[
f(x) = \sum_{z} \widehat{f}(z)\X_z(x) 
\mbox{ }\mbox{ }\mbox{ and } \mbox{ }\mbox{ }
\widehat{f}(z) = \langle f, \X_z\rangle  =\EX_{U_n} f \X_z.
\]

The {\em degree}  of $f\in \L(\Z_2^n)$ 
 is the smallest 
degree of a polynomial  $p\in \cmp[x_1,\ldots,x_n]$ such that 
$p(x) = f(x)$ for all $x\in \{0,1\}^n$.
Equivalently, in terms of the Fourier transform $\widehat{f}$, 
the degree of $f$ is equal to the maximal weight of 
 $z\in \Z_2^n$ such that  $\widehat{f}(z) \neq 0$.

In terms of the characters $\{\X_z\}_z$,  
 we have the following equivalent definition of $k$-wise independence.  
  A probability distribution $\mu$ on $\{0,1\}^n$ is  $k$-wise independent
  iff  
  $\EX_\mu \X_z = 0$   for each nonzero $z\in \{0,1\}^n$  such that $|z|\leq k$. 
  Equivalently, $\mu$ is $k$-wise independent iff $\EX_\mu p = \EX_{U_n} p$ for each
  polynomial $p(x_1,\ldots, x_n) \in \cmp[x_1,\ldots, x_n]$ of degree at most $k$.
  This follows from the fact that the evaluation $f$ of $p$ on $\{0,1\}^n$ has degree at most $k$, hence 
  its Fourier transform $\widehat{f}$ 
  is zero on all frequencies of weight larger than $k$, i.e.,  
  $p(x) = \sum_{z\in \{0,1\}^n: |z|\leq k}\widehat{f}(z) \X_z(x)$, for all $x\in \{0,1\}^n$.

\section{Proof of Lemma \ref{covlpduality}}\label{covlpdualitypms} 
The lemma is restated below for convenience. 
\medskip \\
{\bf Lemma \ref{covlpduality}}~{\em
    Let $1\leq k\leq n$ be integers and $R>0$ a real number. Then   the following are equivalent:
  \begin{itemize}
  \item[I)]  Each $k$-wise independent probability distribution on $\{0,1\}^n$ has  covering radius
    less than $R$
   \item[II)] There exits a polynomial $p(v)\in \R[v]$ of degree at most $k$ such that $\EX_{B_n} p>0$ and 
          $p(w)\leq 0$, for each $w\in [0:n]$ such that $w\geq R$.  
   \end{itemize}}
\smallskip \noindent
First, we note that (I) is equivalent to:
        \begin{itemize}
  \item[I')]  For each $k$-wise independent probability distribution on $\{0,1\}^n$, we have 
            $\mu(\B_n(0; r))\neq 0$, where $r = \ceil{R}-1$. 
   \end{itemize}
  The reason is that    $\mu(\B_n(x; r)) = (\sigma_x \mu)(\B_n(0; r))$,
  where 
  $\sigma_x \mu$  is the translation of $\mu$ by $x$
  (i.e., $(\sigma_x \mu)(y)=\mu(x+y)$). The equivalence between (I) and (I') then follows from the fact that
  if $\mu$ is $k$-wise independent, then so is $\sigma_x \mu$ 
  because $\EX_{\sigma_x\mu} \X_z= \X_z(x) \EX_\mu\X_z$. That is, we may assume 
  without loss of generality that $x=0$.

  The equivalence between (I') and (II) follows from Linear Programming duality.  The use of LP duality in such problems goes back to  Delsarte \cite{Del73}.
  Before going to the LP formulation, it is instructive to directly establish the implication from (II) to (I') by appropriately translating 
  Tiet\"{a}v\"{a}inen's argument to the distributions framework. Assume that (II) holds and let $p$ be such a 
  polynomial.  Consider any  $k$-wise independent distribution $\mu$ on $\{0,1\}^n$.
   Let $I$ be the set of $w\in [0:n]$ such that
  $w<R$, i.e., $w\leq r$,  and let $I^c$ be the complement of $I$ in $[0:n]$. 
  Let $M$ be the maximum value of $p$ in $I$. 
  Since $\EX_{B_n} f>0$ and $f$ is non-positive on $I^c$, $M$ must be positive. Let $p'  = \frac{p}{M}$.
  Thus $p' \leq 1$ on $I$ and $p\leq 0$ on $I^c$, i.e., 
  $p'(w) \leq \delta_I(w)$ for each $w\in [0:n]$, where  $\delta_I$ is the indicator function 
  of $I$ (for each $w\in [0:n]$, 
  $\delta_{I}(w) =1$ if $w\leq r$ and, otherwise,  $\delta_{I}(w) = 0$). Therefore, 
  $$ \mu(\B_n(0; r)) = \EX_{x\sim \mu} \delta_I(|x|) \geq \EX_{x\sim \mu} p'(|x|) = \EX_{x\sim U_n} p'(|x|) =
  \EX_{B_n} p' >0,$$
  where the second equality follows from the fact the $\mu$ is $k$-wise independent and 
  $p'(x_1+\ldots +x_n) \in \R[x_1,\ldots, x_n]$ is a polynomial on the variables $x_1,\ldots, x_n$
  of degree at most $k$.

  Now, we establish the lemma using linear programming duality. Note that the above argument is not enough for our purposes since
  Theorem \ref{mexi} follows from Theorem \ref{mainth} via
  the other implication from (I') to (II).  
  Consider the      linear program 
  \[
        A= \min_{\mu} \mu(\B_n(0; r)),
   \]
  where the minimum is over all $k$-wise independent probability distributions on $\{0,1\}^n$.
  Note that objective function is $\mu(\B_n(0; r))=  \EX_\mu f$, where
  $f$ is the indicator function of $\B_n(0; r)$, i.e, $f(x) = 1$ if $|x|<R$ and $f(x) = 0$ if $|x|\geq R$. 
  The linear constraints are $\mu\geq 0$,   $\sum_x \mu(x)  = 1$, and $\EX_\mu \X_z = 0$ for each nonzero $z\in \{0,1\}^n$ such that
  $|z|\leq k$.

  Taking the dual, we  get
  \[
     B    = \max_{q} \EX_{U_n} q, 
  \]
  where the maximum is over all functions $q: \{0,1\}^n \rightarrow \R$ such that the degree of $q$ is at most $k$, i.e, $\widehat{q}(z) = 0$, for each $z\in \{0,1\}^n$ such that $|z|>k$, and    $q \leq f$ pointwise, i.e.,   $q(x)\leq f(x)$   for each $x\in \{0,1\}^n$.
  See Lemma 5.2.10 in  \cite{Baz03} for the underlying
  duality calculations.

  Since the primal is feasible ($U_n$ is a feasible solution) and   bounded (at least $0$), we get that
  $A = B$. That is, (I') is equivalent to:
  \begin{itemize}
      \item[II')]  There exists   
  $q: \{0,1\}^n \rightarrow \R$ such that the degree of $q$ is at most $k$, $\EX_{U_n} q >0$, and
  $q(x)\leq 0$,   for each $x\in \{0,1\}^n$ such that $|x|\geq R$. 
  \end{itemize}
  Note  that  we  dropped the condition $q(x)\leq 1$, for $|x|< R$, since it follows from appropriately scaling $q$. 
  Thus (II) is the special case of (II') corresponding to the case when $q(x)$ is symmetric, i.e., $q(x)$ depends on the weight $|x|$ of $x$. 
  The fact that (II) and (II')  are equivalent follows  from a 
  classical symmetrization argument. 
  Let $p$ be the symmetric polynomial associated with $q$, i.e.,
  $p(w) = \EX_{x:|x|=w} q(x)$,  for all $w\in [0:n]$.  Thus $\EX_{B_n} p = \EX_{U_n} q>0$ and  
  $p(w) \leq 0$ for each $w\geq R$.      To see why  $p$ has degree at most
 $k$ in $w$, 
  consider the Fourier expansion of $q$: $q(x) = \sum_{|z|\leq k} \widehat{q}(z) \X_z(x)$.
  Thus
  \begin{equation}\label{eqkraw}
    p(w) = \sum_{|z|\leq k} \widehat{q}(z) \EX_{|x|=w}\X_z(x) =
    \sum_{|z|\leq k} \widehat{q}(z) 
    \frac{1}{\binom{n}{w}} K_{w}^{(n)}    (|z|) =  
    \sum_{|z|\leq k} \widehat{q}(z) 
  \frac{1}{\binom{n}{|z|}} K_{|z|}^{(n)}(w), 
  \end{equation}
  where $K_t^{(n)}(w) = \sum_{|z|=t} \X_z(x) =   \sum_{i=0}^t (-1)^i \binom{w}{i}\binom{n-w}{t-i}$
  is the degree-$t$ Krawtchouk polynomial and $x$ is any element of $\{0,1\}^n$ of weight $w$. 
  Note that (\ref{eqkraw}) uses the
  Krawtchouk polynomials identity   $\binom{n}{t} K_w^{(n)}(t) = \binom{n}{w} K_t^{(n)}(w)$ (e.g., see (2.3.15) in \cite{CC97}).


\section{Approximation theory machinery}\label{appmach}

Consider the space $C[-1,1]$  of continuous function on the interval $[-1,1]$ endowed with the max norm: 
\[
  \|f\|_{[-1,1]} = \max_{-1 \leq x \leq 1} |f(x)|.
  \]

\paragraph
    {\bf Lebesgue Constant.}
    Let $X = \{x_i\}_{i=1}^{k+1}$ be an increasing sequence of real points in the interval $[-1,1]$. In what follows, we assume that $k \geq 1$.     
The {\em Lebesgue constant} 
of $X$   is
given by $$\Lambda_k(X) = \max_{p} \| p\|_{[-1,1]},$$  where the maximum is over the choice of a  polynomial $p\in \R[x]$ of degree at most $k$ such that $|p(x_i)|\leq 1$ for $i=1,\ldots, k+1$.

In interpolation theory,  $\Lambda_k(X)$ captures  how good are interpolations on $X$ of functions in $C[-1,1]$ by 
degree-$k$ polynomials  in comparison   to  optimal  degree-$k$ polynomial approximations  with respect the max norm on $[-1,1]$. 
    For our purposes, the above simple equivalent definition is enough. We also need the following estimates of the Lebesgue constant of specific sequences;  e.g., see Section 1.4 in  \cite{MM08}.

{\em Equally-spaced sequences.}
Let  $E^{(k)}$ be the sequence 
of $k+1$ equally-spaced points starting with $-1$ and ending with $1$.
  Then, as $k$ tends to infinity,    $\Lambda_k(E^{(k)}) \sim \frac{2^{k}}{e k \log{k}}$. 
  We also have the bound $    \Lambda_k(E^{(k)}) < \frac{2^{k+3}}{k}$,   which holds for all $k \geq 1$.

  {\em Extended Chebyshev sequences.}  
  The extended  Chebyshev sequence 
  $C^{(k)} = \{c_i\}_{i=1}^{k+1}$ is the increasing sequence 
  given by
  \[
        c_i = -\frac{\cos{(2i-1)\phi_k}}{\cos{\phi_k}}.
  \]
Thus  $c_1  = -1$ and $c_{k+1} =1$. Extended Chebyshev sequences have much better Lebesgue constants than equally-spaced ones. 
  As $k$ tends to infinity, we have the estimate    
  $\Lambda_k(C^{(k)}) \sim \frac{2}{\pi} \log{k}$. 
  We also have the  bound: 
\begin{equation}\label{lebcheb}
  \Lambda_k(C^{(k)}) <  \frac{2}{\pi} \log{(k+1)}  + 0.7213 
  ~~~ \mbox{ for all $k \geq 1$.} 
  \end{equation}
  For our purposes, the fact that $\Lambda_k(C^{(k)}) = O(\log{k})$ is sufficient.

  \paragraph{Bounds outside $[-1,1]$.} We need the following basic
  tool from approximation theory which bounds the absolute value of a polynomial on points outside the interval $[-1,1]$ in terms of its
  max norm on $[-1,1]$ and its degree. 
  \begin{lemma}\label{basapf}    If $p\in \R[x]$ of degree at most $k$,  then for each  real  $x$ such that $|x|>1$,
  \[
  |p(x)| \leq  \| p\|_{[-1,1]} (2|x|)^k
  \]
  \end{lemma}
  Lemma \ref{basapf}  follows from properties of Chebyshev polynomial.
  If $k\geq 0$ is an integer, 
 the $k$'th Chebyshev polynomial
 of the first kind is a degree-$k$ polynomial             $T_k(x)\in \R[x]$ given by 
\[ 
            T_k(x) = \frac{1}{2}\left((x+\sqrt{x^2-1})^k + (x-\sqrt{x^2-1})^k  \right).   
\] 
See   \cite{Riv90} and  \cite{SV14} for a general reference on Chebyshev polynomials.   Lemma  \ref{basapf} is a consequence of
the following basic basic facts about Chebyshev polynomials:  
\begin{itemize}
\item 
  If $p\in \R[x]$ is of degree at most $k$,  then for each  real  $x$ such that $|x|>1$,
  \[
  |p(x)| \leq \| p\|_{[-1,1]} |T_k(x)|.  
  \]
\item If $|x|\geq 1$, then $|T_k(x)|\leq  (2|x|)^k$. This follows immediately from the definition of $T_k$.  
\end{itemize}

\section{Scaling, translation,  and distortion}
\label{std}

We are interested  in integer sequences in the interval $[0: n]$.
Since the Lebesgue constant is invariant under scaling and translations, the
above machinery directly translates from the interval $[-1,1]$ to any interval in $\R$. 
We  introduce in this section the needed 
notations. Then we note that a direct consequence of Markov's theorem is that the Lebesgue constant of a sequence does not significantly
increase after small distortions of its points. 
Distortions will result in this paper  from quantizing  
 real sequences in the real interval interval $[0,n]$ to integer values in the discrete interval $[0: n]$.

Let $X = \{x_i\}_{i=1}^{k+1}$ be a sequence of $k+1$ increasing points in $\R$. Define the {\em Lebesgue constant}  $\Lambda(X)$ of $X$ as
 $$\Lambda(X) = \Lambda_k(\bar{X}),$$ where $\bar{X}=\{\bar{x}_i\}_{i=1}^{k+1}$ is the sequence obtained by 
translating and scaling $X$ so that $\bar{x}_1=-1$  and $\bar{x}_{k+1}=1$.

Define the {\em interval}, {\em center}, and {\em radius} of $X$ by $I(X) = [x_1, x_{k+1}]$, $C(X) = \frac{x_1+x_{k+1}}{2}$, and $R(X) = \frac{x_{k+1}-x_{1}}{2}$, respectively.

If $I$ is a closed real interval and $f$ is continuous, let 
$$
\|f\|_{I} = \max_{x\in I} |f(x)|.
$$
Also define
$$
\|f\|_{X} = \max_{i=1}^{k+1} |f(x_i)|.
$$
Thus
\begin{equation}\label{lambdaX}
  \Lambda(X) = \max\{ \| p\|_{I(X)}~:~ p\in \R[x]  \mbox{ of degree at most } k \mbox { such that } \|p\|_X \leq 1\}. 
\end{equation}
Therefore, using (\ref{lambdaX}) with Lemma \ref{basapf}, we get the following bound. 
\begin{corollary} \label{sctr}  {\bf (Key tool)} Let $p\in \R[x]$  be a polynomial of degree at most $k$ and  let 
  $X$ be a sequence of $k+1$ increasing points in $\R$. 
  Then, for each  real  $x$ outside $I(X)$, 
  \[
  |p(x)| \leq  \| p\|_{X} \Lambda(X) \left|\frac{2(|x|-C(X))}{R(X)}\right|^k. 
  \]
\end{corollary}
Corollary \ref{sctr} is the key tool in the proof of Theorem \ref{mainth}. 
To handle distortions, we need Markov's theorem. 
\begin{lemma} [Markov's theorem; see \cite{Che66}] If $p\in \R[x]$ is a degree $k$ polynomial, consider the derivative $p'$ of $p$.
  Then $\|p'\|_{[-1,1]} \leq k^2  \|p\|_{[-1,1]}$. 
  \end{lemma}
\begin{corollary}\label{distor} {\bf (Distortion)}
  Let $X = \{x_i\}_{i=1}^{k+1}$ and $X' = \{x_i'\}_{i=1}^{k+1}$ be two increasing  sequences of points in $\R$.
Let $\gamma >0$  be such that $\gamma k^2 \Lambda(X) < 1$.  
Assume that $|x_i' - x_i| \leq \gamma R(X)$,
for $i=1,\ldots, k+1$,  and  that 
  $I(X') \subset I(X)$ (i.e,   $x_1 \leq x_1' \leq x_{k+1}$ and $x_1 \leq x_{k+1}' \leq x_{k+1}$). Then 
  \[
  \Lambda(X') \leq \frac{\Lambda(X)}{1- \gamma k^2 \Lambda(X)}
  \]
  \end{corollary}
    {\bf Proof:}  Since the Lebesgue constant is  invariant under scaling and translation, assume without loss of generality that
    $I(X) = [-1, 1]$, i.e., $C(X)= 0$ and $R(X)=1$. Let $p\in \R[x]$ be  of degree at most $k$ such that  $\|p\|_{X'}\leq 1$. 
    For any $1 \leq i\leq k+1$, we have $|p(x_i)-p(x_i')| \leq \gamma\| p'\|_{[-1,1]}$ since   $x'_i\in I(X) = [-1,1]$. Applying Markov's theorem, we get
    $|p(x_i)-p(x_i')| \leq \gamma k^2 \| p\|_{[-1,1]}
    $. Hence $|p(x_i)| \leq |p(x_i')| +\gamma k^2\| p\|_{[-1,1]}$.
   It follows that
   $$\|p\|_{X} \leq  \|p\|_{X'} + \gamma k^2\| p\|_{[-1,1]}
   \leq 1 + \gamma k^2\Lambda(X)\| p\|_{X}. 
   $$
   Therefore
   $$
   \|p\|_{X} \leq  \frac{1}{1- \gamma k^2 \Lambda(X) }.
   $$
   Hence 
   $$
   \|p\|_{I(X')} \leq \|p\|_{[-1,1]}
   \leq  \frac{\Lambda(X)}{1- \gamma k^2 \Lambda(X) }, 
   $$
   where the first inequality holds because 
   $I(X')\subset I(X) = [-1,1]$. 
    \finito

\section{Proof Technique}\label{prtech}
Consider the following linear program.
\begin{definition}
  If $n,k\geq 1$ are integers and
  $\Delta >0$ is a real number such that $\Delta\leq n/2$, let
  \[
         \E_n(k,\Delta) = \max \EX_{B_n} f,  
         \]
  where the maximum is over all  polynomials $f \in \R[x]$ such that the degree of $f$ is at most $k$ and 
  $f(w)\leq 0$, for each integer $w\in [0:n]$ such that  $|w - n/2|\leq \Delta$.
\end{definition}
Note that, by setting $f$ to the identically zero polynomial, we get    $\E_n(k,\Delta) \geq 0$.
\begin{definition}
  If $n,k\geq 1$ are integers,
  let
  $\Delta^*_n(k)$ be the minimum value of
  $\Delta>0$ such that $\E_n(k,\Delta)=0$. 
  \end{definition}
In the above terms terms, Theorem \ref{mainth} can be restated as follows.
\medskip \\
{\bf Theorem \ref{mainth}}~{\em
  There exist   absolute constants $k_0,n_0>0$ such that 
  for each  integer $n\geq n_0$ and each  integer $k$ satisfying 
  $k_0 \leq k \leq \frac{n^{1/3}}{\log^2{n}}$, we have
  have $\E_n( k, \sqrt{kn})= 0$, or equivalently,   
$\Delta^*_n(k) \leq \sqrt{kn}$.} 
\smallskip\\
We will actually prove a slightly stronger statement; we will show that  $\Delta^*_n(k) \leq \sqrt{\al kn}$, 
    where $\al =0.93$.   
    This bound is asymptotically tight up to a factor less than $2$; 
        it follows from  Tiet\"{a}v\"{a}inen's bound  (Lemma \ref{trl90}) 
        that if  $1 \leq k \leq n-1$ are integers, 
        then 
\[
  \left\{\begin{array}{ll}
    \Delta_n^*(k) > \sqrt{s(n-s)} - s^{1/6} \sqrt{n-s} & \mbox{if $k=2s-1$ is odd}\\
    \Delta_n^*(k) > \sqrt{s(n-1-s)} - s^{1/6} \sqrt{n-1-s} + \frac{1}{2} & \mbox{if $k=2s$ is even.} 
  \end{array}\right.
  \]

To illustrate the technique,  we assume below that $\Delta>0$ is any number such that $\Delta \leq n/2$. Given $k$, we would like to make $\Delta$ as small as possible while guaranteeing that $\E_n(k,\Delta)=0$.  
We illustrate  in this section how to reduce  the task of showing that $\E_n(k, \Delta) = 0$  to that of constructing a sequence with appropriate parameters (Lemma \ref{lemotrseq}).
The outline of the rest of the proof  is  Section \ref{prout}.

Let $L$ be the set integers $w \in [0:n]$ such that  that  $|w-\frac{n}{2}|\leq \Delta $ and let $L^c$ be the complement of $L$ in $[0:n]$. 
Let $f(x)\in \R[x]$ be a polynomial of degree at most $k$, where $k\geq 1$ is an integer. 
Assume that $f(w)\leq 0$, for each  $w\in L$.  
We want to show that $\EX_{B_n} f \leq  0$ if $k$ is small enough compared to $\Delta$.

Let  $W= \{w_{i}\}_{i=1}^{k+1}$ be  a length-$(k+1)$ {\em integer} 
sequence of  increasing  points contained in the interval $L$ and centered at $n/2$. 
  By Corollary \ref{sctr}, for each $w \in L^c$, 
\[
    |f(w)| \leq  \| f\|_{W} \Lambda(W) \left|\frac{2(w-n/2)}{R(W)}\right|^k.
    \]
Let $w^*$ be the point in
$W$ which maximizes $\|f\|_W$, i.e., $w^* = w_{i^*}$, where $i^*$ is such that $|f( w_{i^*} )| = \|f\|_{W}$.  Since $f\leq 0$ on $L$,
$f( w^* ) = -\|f\|_{W}$.  
Note that  $\|f\|_{W} \neq 0$ unless $f$ is identically zero 
since the degree of $f$ is at most $k$ and    $W$ has $k+1$ points. They key is to try to  use the  point $w^*$ to bound $\EX_{B_n} f$ as follows. 
We have
\[ 
\EX_{B_n} f
= \sum_{w=0}^n B_n(w) f(w)
  \leq \sum_{w\in L^c} B_n(w) |f(w)| -  \|f\|_{W} B_n(w^*). 
  \]
  As we don't have information about the position of $w^*$ in $W$, we use the following  bound 
\[
  B_n(w^*) \geq B_n\left({\frac{n}{2} + R(W)}\right),  
  \]
  which  follows from the fact that the binomial distribution
  $B_n$ is bell shaped around $n/2$. 
  Note that, even if $n$ is odd,  $\frac{n}{2}+R(W)$ is an integer since $W$ is an integer sequence 
  centered at $n/2$ with radius $R(W)$. 
  It follows that 
  \begin{eqnarray*}
  \EX_{B_n} f 
  &\leq&  
   \|f\|_{W} \left(  \Lambda(W)
  \sum_{w\in L^c} B_n(w) \left|\frac{2(w-n/2)}{R(W)}\right|^k -
  B_n\left(\frac{n}{2} + R(W)\right)   \right)\\
  &=&
     \|f\|_{W} \left( 2 \Lambda(W)
  \sum_{w> \frac{n}{2}+\Delta} B_n(w) \left(\frac{2(w-n/2)}{R(W)}\right)^k -
  B_n\left(\frac{n}{2} + R(W)\right)   \right). 
  \end{eqnarray*}
  In summary, we get the following: 
  \begin{lemma}\label{lemoneseq}
    {\bf (One-sequence approach)}
    Let $n,k \geq 1$ be integers and let  $\Delta>0$ be a real number such that $\Delta \leq n/2$.  
Let $L$ be the set integers $w \in [0:n]$ such that  that  $|w-\frac{n}{2}|\leq \Delta $. 
Let $W$ be  a length-$(k+1)$  integer
sequence of  increasing  points contained in the interval $L$ and centered at $n/2$. 
Let 
\[
\nu = 2\Lambda(W) \sum_{w> \frac{n}{2}+\Delta} B_n(w)
\left(\frac{2(w-n/2)}{R(W)}\right)^k.  
  \]
  If $\nu \leq  B_n\left(\frac{n}{2} + R(W)\right)$,  then $\E_n(k,\Delta) = 0$. 
  \end{lemma}
    
  \paragraph       {\bf  Limitations of the one-sequence approach.}
Consider the setup when $\Delta = \sqrt{\al kn}$, where $\al>0$ is any constant. 
  To motivate the translated sequences approach explained below, 
  we note below that the one-sequence approach is not useful if $k$ is small.
  It can be used to establish  Theorem \ref{mainth} for $k = w(\log{n})$,
  but it it fails for smaller values of $k$. Namely, for all constants $\al>0$, it fails 
  to show that $\E_n(k, \sqrt{\al kn})=0$  if $k = o(\log{n})$.

  Assume that $k  = o(\log{n})$.
  Since $W$ is contained in $L$,  $R(W) \leq \Delta$.
  Using the loose lower bounds  
  $\frac{2(w-n/2)}{R(W)} > \frac{2 \Delta }{R(W)} \geq 2 > 1$,
  for  $w  > \frac{n}{2}+\Delta$, and  $\Lambda(W) \geq 1 > \frac{1}{2}$, we get 
  \[
  \nu  >   \sum_{w> \frac{n}{2}+ \sqrt{\al kn}} B_n(w)
  \geq   \sqrt{\al k n}   
  B_n\left( \ceil{ \frac{n}{2} + 2 \sqrt{\al k n} } \right)
  = \Omega\left( \frac{\sqrt{k n}}{ \sqrt{n} } e^{-8\al k} \right)
    = \Omega\left( \sqrt{k}  e^{-8\al k} \right), 
\]
  via 
   de Moivre-Laplace normal approximation of the binomial (see Theorem
   \ref{demlt}). 
On the other hand, we have  
     $$
  B_n\left(\frac{n}{2}+R(W)\right) \leq
  B_n\left( \floor{ \frac{n}{2}  } \right)
  = \Theta\left( \frac{1}{\sqrt{n}} \right).
  $$
  Thus, to conclude that 
  $\nu \leq  B_n\left(\frac{n}{2} + R(W)\right)$, 
  we need $k$ to be at least $\Omega(\log{n})$ to compensate for the
  $\frac{1}{\sqrt{n}}$ term. 
   
  The one-sequence approach exhibits one point $w^*$ in the sequence $W$ on which
  $f$ is negative (assuming that $f$ is not identically zero).  
  To  resolve the   $\frac{1}{\sqrt{n}}$  issue, we will use 
  multiple disjoint sequences and exhibit one point in each sequence on which 
  $f$ is negative.
  The sequences will be translates of $W$. To guarantee that they are disjoint, their number is limited by the minimum distance $t$ between  consecutive points in $W$. Eventually, we will overcome the   $\frac{1}{\sqrt{n}}$ term
  by using a sequence with
  $t = \Omega(\frac{\sqrt{n}}{k^{O(1)}})$.

\begin{lemma}\label{lemotrseq}
       {\bf (Translated  sequences approach)}
Let $n,k \geq 1$  integers and let  $\Delta>0$ be a real number such that $\Delta \leq n/2$.  
Let $W= \{w_{i}\}_{i=1}^{k+1}$ be 
a length-$(k+1)$ integer sequence of  increasing  points centered at
$n/2$.  
  Let $t$ be the minimum distance  between two consecutive points in $W$,
  i.e., $t = \min_{i=1}^{k} w_{i+1}-w_i$, and 
 let $p = \ceil{\frac{t-1}{2}}.$ 
Assume that $R(W)+p\leq \Delta$.
Let 
\[
\nu =  2\Lambda(W)   \sum_{w>\frac{n}{2}+\Delta} B_n(w) \left(\frac{2(w-n/2)+2p}{R(W)}\right)^k. 
\]
If $\nu   \leq    t B_n\left(\frac{n}{2} + R(W)+ p\right)$,  then $\E_n(k,\Delta) = 0$.
  \end{lemma}
{\bf Proof:}  
Consider the $t$ translated sequences 
$W_0,\ldots, W_{t-1}$ , where for $s = 0,\ldots, t-1$, 
$W_s = \{w_{s,i}\}_{i=1}^{k+1}$ and
$w_{s,i} = w_{i+s-p}$. 
  Thus $W = W_{p}$.  By the definition of $t$, the sequences $W_0,\ldots, W_{t-1}$ 
  are disjoint, i.e., $w_{s,i} \neq w_{s',i'}$ if $(s,i)\neq (s',i')$.
  Moreover, for each $s$, $R(W_s) = R(W)$, $\Lambda(W_s)=\Lambda(W)$, 
  and, since  $W$ is centered at $n/2$,  $C(W_s) = n/2 +  s -p$. Thus
\begin{equation}\label{blabla1}
  |C(W_s)-n/2|\leq p.
  \end{equation}
  As above,  let $L$ be the set integers $w \in [0:n]$ such that  that  $|w-\frac{n}{2}|\leq \Delta $ and let $L^c$ be the complement of $L$ in $[0:n]$.   
  Since $R(W)+p\leq \Delta$, 
  each $W_s$ is contained in $L$.

By Corollary \ref{sctr}, for each $w \in L^c$ and for each $0 \leq s \leq t-1$, 
\[
    |f(w)| \leq  \| f\|_{W_s} \Lambda(W) \left|\frac{2(w-C(W_s))}{R(W)}\right|^k.
    \]
    Averaging over $s$, we get
    \begin{equation}\label{avleb}
      |f(w)| \leq  \frac{\Lambda(W)}{t}\sum_{s=0}^{t-1}\| f\|_{W_s} 
      \left|\frac{2(w-C(W_s))}{R(W)}\right|^k.
    \end{equation}
    Now we argue as above on each $W_s$. 
    For each $s$, 
let $w^*_s$ be the point in
$W_s$ which maximizes $\|f\|_{W_s}$, i.e., $w^*_s = w_{s,i^*}$, where $i^*$ is such that $|f( w_{s,i^*} )| = \|f\|_{W_s}$.  Since $f\leq 0$ on $L$,
$f( w_{s}^* ) = -\|f\|_{W_s}$.  Here again,  $\|f\|_{W_s} \neq 0$ if $f$ is not identically zero.  
The key is to  use the integer points  $\{w_s^*\}_{s=0}^{t-1}$ to bound $\EX_{B_n} f$. Note that
$w_0^*, \ldots, w_{t-1}^*$ are distinct points  contained in $L$ since the sequences $W_0, \ldots, W_{t-1}$ are disjoint and  contained in $L$.  
Therefore,
\begin{equation}\label{blabla2}
\EX_{B_n} f
= \sum_{w=0}^n B_n(w) f(w)
  \leq \sum_{w\in L^c} B_n(w) |f(w)| - \sum_{s=0}^{t-1} \|f\|_{W_s} B_n(w_s^*). 
 \end{equation}
  We have
  $$\left|w_s^*-\frac{n}{2}\right|
  \leq \left|w_s-C(W_s)\right| + \left|C(W_s)-\frac{n}{2}\right|     \leq R(W_s) +
  \left|C(W_s)-\frac{n}{2}\right|.$$ Using 
 (\ref{blabla1}) and the fact that
  $R(W_s)=R(W)$, we obtain 
  $
\left|w_s^*-\frac{n}{2}\right|
\leq R(W) + p$, and hence   
\begin{equation}\label{blabla3}
B_n(w_s^*) \geq  B_n\left(\frac{n}{2} + R(W)+ p\right). \end{equation}
As before, note that, even of $n$ is odd,  $\frac{n}{2}+R(W)$ is an integer since $W$ is an integer sequence centered at $n/2$.  If follows also from
(\ref{blabla1}) that 
\begin{equation}\label{blabla4}
  |w-C(W_s)| \leq \left|w-\frac{n}{2}\right| +
  \left|C(W_s)-\frac{n}{2}\right| \leq
  \left|w-\frac{n}{2}\right| + p. 
\end{equation}
Therefore, 
by  replacing (\ref{avleb})  and (\ref{blabla3}) 
in (\ref{blabla2}),  using
(\ref{blabla4}),  
and then interchanging the summations, we get
  \begin{eqnarray*}
&& \EX_{B_n} f 
  \leq
  \frac{1}{t}\sum_{s=0}^{t-1} \|f\|_{W_s} \left(  \Lambda(W)
  \sum_{w\in L^c} B_n(w) \left|\frac{2\left(  \left|w-\frac{n}{2}\right| + p\right)}{R(W)}\right|^k -
  t B_n\left(\frac{n}{2} + R(W)+ p\right)   \right)\\
  &&~ =  
   \left( \frac{1}{t}\sum_{s=0}^{t-1} \|f\|_{W_s} \right) \left(  2\Lambda(W)
  \sum_{w>\frac{n}{2}+\Delta} B_n(w) \left(\frac{2(w-n/2+p)}{R(W)}\right)^k -
  t B_n\left(\frac{n}{2} + R(W)+ p\right)   \right). 
  \end{eqnarray*}
  \finito

\subsection{Discussion and proof outline} \label{prout}
Lemma \ref{lemotrseq} reduces the problem of showing that  
$\E_n(k,\Delta) = 0$ to that of constructing the sequence $W$.
The relevant parameters of $W$ are
its radius $R(W)$, its Lebesgue constant $\Lambda(W)$, and its
minimum distance $t$. We need 
$t$ to be large and    $\Lambda(W)$ small. We also need to optimize on
$R(W)$ since 
increasing $R(W)$ decreases both $\nu$ and
$t B_n\left(\frac{n}{2} + R(W)+ p\right)$. 

In the next section, we will construct $W$ by starting with  
translated and a scaled Chebyshev sequence $X$ and quantizing its points to
integer values. 
We will see that,  for a suitable choice of   parameters, the effect of quantizing is negligible as it increases its Lebesgue constant  by at most a factor of $2$. This follows from Markov's theorem via  Corollary 
\ref{distor}.
For $\Delta = \Theta(\sqrt{kn})$,
$R(W) = \Theta(\Delta)$, and $k \leq \frac{n^{1/3}}{\log{n}}$,
we will get
$t= \Theta(\frac{\sqrt{n}}{k^{3/2}})$. For such values,
$B_n\left(\frac{n}{2} + R(W)+ p\right)= \Theta(\frac{1}{\sqrt{n}}e^{-\Theta(k)})$. Hence multiplying $B_n\left(\frac{n}{2} + R(W)+ p\right)$ by $t$ cancels out
the $\frac{1}{\sqrt{n}}$ term and replaces it with  
a  $O(\frac{1}{k^{3/2}})$ term, which barely affects the exponent.

We conclude the proof of Theorem \ref{mainth} in Section \ref{concprf}
by optimizing on $R(W)$ and estimating $\nu$ using
Moivre-Laplace normal approximation of the binomial and  Hoeffding's inequality.

\subsection{Note on equally-spaced sequences} 
Note that the largest possible values of $t$ 
is around $\frac{2R(W)}{k} =  \Theta(\frac{\sqrt{n}}{\sqrt{k}})$ and it is achieved by a sequence of equally-spaced points.  Compared to Chebyshev sequences, 
the gain is negligible since the effect of 
$\frac{1}{k^{3/2}}$ vanishes asymptotically.
The issue with equally-spaced   sequences is that 
$\Lambda(W)$ is exponential in $k$  and namely  around $\frac{2^k}{e k \log{k}}$ (see Section  \ref{appmach}).  Ignoring non-exponential terms,
the effect of using an
equally-spaced sequence boils down to turning the  $2(w-n/2)$ term in the expression of $\nu$ into  $4(w-n/2)$. This    increases the  upper  bound
on $\Delta_n^*(k)$ by a constant factor. 
It can be shown that equally-spaced  sequences lead to  a weaker version of Theorem \ref{mainth} and  namely  that $\Delta_n^*( k)\leq  \sqrt{(1.43) k n}$,  
if $k\leq \frac{n^{1/3}}{\log^2{n}}$ and $k$ and $n$ are sufficiently large.

\section{Quantized Chebyshev sequences}\label{chebybound}

Lemma \ref{implem} below summarizes the parameters
of the sequence $W$ on which Lemma \ref{lemotrseq} will be applied. The sequence is a  quantized version of a scaled and translated  Chebyshev sequence.   
Let $\Delta = \sqrt{\al kn}$,  where $k \leq \frac{n^{1/2}}{\log{n}}$ and
  $\al > 0$ is  a constant (which, as previously mentioned, will be eventually set to $\al =  0.93$).  
The sequence  consists  of integer points. It is  centered at $n/2$  and its   radius is 
$R(W) = \floor{\beta \Delta +1}$, where $0< \beta < 1$ is a constant which we will optimize on in Section \ref{concprf}.
Eventually, we will set 
$\beta = 0.5204$.  
The smallest distance $t$ between consecutive points in $W$ is in the order of  $t = \Omega(\frac{\sqrt{n}}{k^{3/2}})$.
We also have 
$R(W)+p\leq (\beta + \epsilon)\Delta $ and ${2p}\leq \e\Delta$, where $\epsilon>0$ will be set to a  sufficiently small value in Section \ref{concprf}.  We will use $\e$  in Section \ref{concprf}
to   handle the term  $\frac{2p}{R(W)} = \Theta(\frac{1}{k^2})$ and the
offset $p$ added to the radius of $W$ in Lemma \ref{lemotrseq}. 
Eventually, we will set  $\epsilon = 0.004$.   
The Lebesgue constant of the quantized Chebyshev sequence is at most twice that of the original 
Chebyshev sequence.

\begin{lemma} \label{implem}
  Let   $\al, \beta, \epsilon>0$ be positive constants such that $\beta + \epsilon < 1$. 
  Then there exist   $k_1>0$ and $n_1>0$, depending on $\al, \beta,$ and $\e$,  such that 
  for each integer $n \geq n_1$  and each integer $k$ satisfying 
  $k_1 \leq k \leq \frac{n^{1/3}}{\log{n}}  $, the following holds. 

  Let $\Delta = \sqrt{\al kn}$.   Then there exists  a length-$(k+1)$ integer sequence $W$ of  increasing  points centered at $n/2$ such that: 
\begin{itemize}
  \item [a)]
The minimum distance $t$  between consecutive points in $W$ is at least $b(k) \sqrt{n}$, where 
$b(k) = \frac{3\beta }{2}\left(\frac{\pi}{2(k +1)}\right)^2 \sqrt{\al k}$.
\item[b)]  $R(W)\geq \beta \Delta$
  \item[c)]    $R(W)+p\leq (\beta + \epsilon)\Delta$, where $p = \ceil{\frac{t-1}{2}}$ 
\item[d)]  ${2p}\leq \epsilon \Delta$
\item[e)] $\Lambda(W) \leq  \frac{4}{\pi}\log{(k +1)} +2$
\item[f)] $R(W)+p \leq \Delta \leq \frac{n}{2}$.  
\end{itemize}  
\end{lemma}
\paragraph{Proof}
Let $n,k \geq 1$  
and assume that:
     \begin{eqnarray}
    \sqrt{\al k n}  &\leq& \frac{n}{2}  \ \label{qaz02}\\
                \frac{4}{3 \beta \sqrt{\al k}} \left(\frac  {2(k+1)}{\pi }\right)^2 &\leq & \sqrt{ n}  \label{qaz1}\\
                \frac{2k^{3/2}}{\beta\sqrt{\al}}                 \left( \frac{2}{\pi} \log{(k+1)}  + 1\right) &\leq& \sqrt{  n}   \label{qaz2}\\                
                4\left(\frac{\pi}{2(k+1) }\right)^2
\left(\beta   +\frac{1}{\sqrt{\al kn}}\right)
                + \frac{3}{ \sqrt{\al kn}} &\leq& {\epsilon}.\label{qaz4}
     \end{eqnarray}
     We will verify the lemma under the above assumption. Then we show that they hold for $k$ and $n$ large enough if $k\leq \frac{n^{1/3}}{\log{n}}$.

     Note first that since $\beta + \epsilon<1$,  (f) follows trivially from (c) 
     and  (\ref{qaz02}).

Recall from  Section \ref{appmach} the extended Chebyshev sequence    $C= \{c_i\}_{i=1}^{k+1}$, where  
  \[
        c_i = -\frac{\cos{(2i-1)\phi_k}}{\cos{\phi_k}} ~~~~ \mbox{for $i=1,\ldots, k+1$} 
  \]
and $\phi_k = \frac{\pi}{2(k+1)}$. 
Thus  $c_1  = -1$ and $c_{k+1} =1$. By  scaling and translation,  map $C$ into
a real  sequence $X = \{ x_i\}_{i=1}^{k+1}$ centered at $n/2$  with radius $R({X}) = \beta \sqrt{\al kn}+1$. 
That is, ${x}_i = (\beta \sqrt{\al kn}+1)x_i+n/2$. Quantize ${X}$ to construct an an integer sequence $W = \{w_i\}_{i=1}^{k+1}$
centered at $n/2$ as follows.  To make sure that $W$ is centered at $n/2$ and that $I(W)\subset I(X)$,  let
$w_1 = \ceil{{x_1}}$ and   
$w_{k+1} = \floor{{x}_{k+1}}$. For $i = 2, \ldots, k$,   set $w_i$ arbitrarily to  $\floor{{x_i}}$ or $\ceil{{x_i}}$. Note that the condition $I(W) \subset I(X)$ is needed by Corollary  \ref{distor}.
Thus the radius of $W$ is 
$$R(W) = \floor{\beta \sqrt{\al kn}+1} \geq \beta \sqrt{\al kn}, $$
which proves  (b).  Moreover,  $|w_i - {x}_i| \leq 1 =  \gamma R({X})$, where $\gamma = \frac{1}{R(X)} \leq  \frac{1}{\beta \sqrt{\al kn}}$. 

First, we need  to verify that the distortion did not collide points in $W$,  and hence  the length of $W$ is 
 equal to the length $k+1$ of $C$.
  The minimum distance between  points in $C$ is $t_C = c_2 - c_1 = \frac{\cos{\phi_k} - \cos{3\phi_k}}{\cos{\phi_k}}$.
 We have the bounds $3 \phi_k^2  \leq t_C \leq 4 \phi_k^2$, which hold for any $0\leq \phi_k \leq \pi/4$. Thus  
 the minimum distance $t$ between  consecutive points in $W$ is satisfies $\underline{t}\leq t \leq \bar{t}$, where 
\[
\underline{t} = 3\beta \left(\frac{\pi}{2(k+1)}\right)^2  \sqrt{\al kn} -2  ~~~ \mbox{and} ~~~ 
\bar{t}=  4\left(\frac{\pi}{2(k+1) }\right)^2
(\beta\sqrt{\al kn}+1) +2.
\]
Condition (\ref{qaz1}) is equivalent to
\begin{equation}\label{mess281}
3\beta \left(\frac{\pi}{2(k+1)}\right)^2 \sqrt{   \al kn} \geq 4, 
\end{equation}
hence $\underline{t } \geq 2$. Therefore, $\underline{t}>0$, and   hence the points in $W$ are distinct.

{\em Proof of (a).} We have 
   $$
t \geq \underline{t} \geq \underline{t} - \left(\frac{3\beta }{2}\left(\frac{\pi}{2(k +1)}\right)^2 \sqrt{\al kn}-2\right) =  \frac{3\beta }{2}\left(\frac{\pi}{2(k +1)}\right)^2 \sqrt{\al kn}
= b(k) \sqrt{n}
,$$
where the second inequality follows from 
(\ref{mess281}).

{\em Proof of (c).} 
We have $R(W) \leq \beta \sqrt{\al k n} + 1$ and $p = \ceil{\frac{{t}-1}{2}} \leq \bar{t}$, hence
$R(W) + p
\leq  \beta \sqrt{ \al k n} +1+\bar{t}$. Therefore, 
\begin{equation*}\label{mess78}
R(W) + p
\leq 
\beta \sqrt{\al kn }
+4\left(\frac{\pi}{2(k+1) }\right)^2(\beta \sqrt{\al  kn}+1) +3\leq (\beta+\epsilon)\sqrt{\al kn}, 
\end{equation*}
where the last inequality is equivalent to  condition  (\ref{qaz4}).

{\em Proof of (d).} 
   We have
   \[
   \frac{2p}{\sqrt{\al kn}} \leq \frac{\bar{t}+1}{\sqrt{\al k n}} =   4\left(\frac{\pi}{2(k+1) }\right)^2
   \left(\beta   +\frac{1}{\sqrt{\al kn}}\right)
   + \frac{3}{ \sqrt{\al kn}} \leq {\epsilon}, 
   \]
   where the last inequality is condition   (\ref{qaz4}).

{\em Proof of (e).}
Recall from Section \ref{appmach} that 
\[
\Lambda(X) = \Lambda(C) <  \frac{2}{\pi} \log{(k+1)}  + 0.7213 < \frac{2}{\pi} \log{(k+1)}  + 1.  
\]
Invoking  Corollary  \ref{distor}, we get  
  \[
  \Lambda(W) \leq \frac{\Lambda({X})}         {1- \gamma k^2 \Lambda(X)} \leq 2 {\Lambda({X})} 
  \]
  if $\gamma k^2\Lambda({X})\leq \frac{1}{2}$.   Since  $\gamma \leq \frac{1}{\beta \sqrt{\al kn}}$, this condition  follows from 
  \[
          2k^2 \left(\frac{2}{\pi} \log{(k+1)} +1 \right)\leq \beta \sqrt{ \al k n},
  \]
  which is equivalent to condition   (\ref{qaz2}).

{\em Asymptotics.}
It remains to show that for each constants $\alpha,\beta, \epsilon>0$,  there exist 
  $k_1>0$ and $n_1>0$ such that  conditions      (\ref{qaz02}),  (\ref{qaz1}), (\ref{qaz2}), and  (\ref{qaz4}) hold 
  for each $n \geq n_1$  and each  $k$ satisfying 
  $k_1 \leq k \leq \frac{n^{1/3}}{\log{n}}  $.

  The claim is straight forward for condition  (\ref{qaz4}) and it does not require a relation between $k$ and $n$.
  Condition (\ref{qaz02}) is equivalent to $k
  \leq \frac{n}{4\al }$, which holds, for $n$ large enough,  since   $k \leq \frac{n^{1/3}}{\log{n}}$.      
     To verify  (\ref{qaz1}), let $k_2$ be large enough so that
     $\frac{(k+1)^2}{\sqrt{k}}$     is increasing in $k$ for $k \geq k_2$. 
Thus, for $k\geq k_2$,
\[
\frac{4}{3 \beta \sqrt{\al k}} \left(\frac  {2(k+1)}{\pi }\right)^2 \leq
\frac{16}{3 \beta \sqrt{\al}\pi^2}\left(k^{3/4}+1\right)^2 \leq
\frac{16}{3 \beta \sqrt{\al}\pi^2}\left(\frac{n^{1/4}}{\log^{3/4}{n}}+1\right)^2
\leq \sqrt{ n }, 
\]
where the last inequality holds 
for sufficiently large $n$.  To see why the  claim holds for (\ref{qaz2}), note that 
\[
\frac{2k^{3/2}}{\beta\sqrt{\al}}                 \left( \frac{2}{\pi} \log{(k+1)}  + 1\right)
\leq
\frac{2}{\beta\sqrt{\al}} \frac{n^{1/2}}{\log^{3/2}{n}} \left( \frac{2}{\pi} \log{(k+1)}  + 1\right) \leq \sqrt{  n},  
\]
where the last inequality holds for $n$ large enough. 

  \finito

\section{Putting things together}\label{concprf}
Let $\al >0$ be a constant. We will show that for
$\al =  0.93$,
there exist   absolute constants $k_0,n_0>0$ such that for all integers $n\geq n_0$ and   $k$ satisfying 
  $k_0 \leq k \leq \frac{n^{1/3}}{\log^2{n}}$, we have 
$\E_n(k,\sqrt{\al kn}) = 0$.

Let $\beta ,\epsilon>0$ be real number such that $\beta+\epsilon < 1$. Assume that $k$ and $n$ are sufficiently large so that  Lemma  \ref{implem} is applicable.
Note that the condition $k \leq \frac{n^{1/3}}{\log^2{n}}$ is stronger than the condition $k \leq \frac{n^{1/3}}{\log{n}}$ required by Lemma  \ref{implem} (the stronger condition is needed by
Lemma \ref{eee5} below).

Let $\Delta =  \sqrt{\al kn}$.  By lemma \ref{implem},  there exists  a length-$(k+1)$  integer sequence $W$ of  increasing  points centered at $n/2$ such that: 
\begin{itemize}
  \item [a)]
The minimum distance $t$  between any two consecutive points in $W$ is at least $b(k) \sqrt{n}$, where 
$b(k) = \frac{3\beta }{2}\left(\frac{\pi}{2(k +1)}\right)^2 \sqrt{\al k}$.
\item[b)]
  $R(W)\geq \beta \sqrt{\al kn}$
  \item[c)]
  $R(W)+p\leq (\beta + \epsilon)\sqrt{\al kn}$, where $p = \ceil{\frac{t-1}{2}}$ 
\item[d)]  ${2p}\leq \epsilon \sqrt{\al kn}$
\item[e)] $\Lambda(W) \leq  \frac{4}{\pi}\log{(k +1)} +2$
\item[f)] $R(W)+p \leq \Delta \leq \frac{n}{2}$.  
\end{itemize}
Applying Lemma \ref{lemotrseq} to $W$, we get that
\begin{equation}\label{thmainconc}
 \E_n(k,\sqrt{\al kn}) = 0
  \end{equation}
if 
\begin{equation}\label{shmuck}
  \nu   \leq    t B_n\left(\frac{n}{2} + R(W)+ p\right),
  \end{equation}
where 
\[
\nu =  2\Lambda(W)   \sum_{w>\frac{n}{2}+\Delta} B_n(w) \left(\frac{2(w-n/2)+2p}{R(W)}\right)^k.
\] 
It follows from (a), (b), (c), (d), and   (e) 
that
\[
t B_n\left(\frac{n}{2} + R(W)+ p\right) \geq b(k) \sqrt{n} B_n\left(\floor{ n/2+ (\beta+\epsilon) \sqrt{\al k n}}\right) 
\]
and
\[
\nu \leq a(k) \sum_{w>\frac{n}{2}+\sqrt{\al kn}} B_n(w) \left(\frac{2(w-n/2) + \epsilon \sqrt{\al kn}}{\beta \sqrt{\al k n}}  \right)^k,  
\]
where $a(k) =  \frac{8}{\pi} \log{(k +1)}  + 4$.

\begin{lemma}\label{eee5}
  Let $\al, \beta>0$ and $\e\geq 0$  be constants   such that:   
  \begin{eqnarray}
&&  \frac{2+\epsilon}{{\beta}} \geq \sqrt{e} \label{messqaz0}\\
    &&  \al > \frac{1}{2}\log\left(\frac{2+\epsilon}{{\beta}}    \right). \label{messqaz}
\end{eqnarray}
  There are constants $n_1,k_1>0$,  depending only on
  $\al$ and $\beta$ and $\epsilon$,
such that for all $n\geq n_1$ and all $k$ satisfying $k_1 \leq k \leq \frac{n^{1/3}}{\log^2{n}}$, we have 
\begin{itemize}
  \item[a)]
$  B_n\left(\floor{ n/2+ (\beta+\epsilon) \sqrt{\al k n}}\right) \geq  \frac{1}{\sqrt{2 n}}e^{-2(\beta + \epsilon)^2 \al k }$

\item[b)]
$\sum_{w>\frac{n}{2}+\sqrt{\al kn}} B_n(w) \left(\frac{2(w-n/2)+\epsilon \sqrt{\al kn}}{\beta\sqrt{\al  k n}}  \right)^k 
  \leq (1+\sqrt{\al k})  e^{-ck}$, where
  $c  =  2\left(\al- \frac{1}{2}\log\left(\frac{2+\epsilon}{{\beta}}\right)\right)$.  
\end{itemize}
{\em Note: } Conditions (\ref{messqaz0}) and (\ref{messqaz}) are not needed in (a). 
\end{lemma}
The proof of Lemma \ref{eee5} is in Section \ref{eee5p}.

Therefore to make sure that (\ref{shmuck})
holds we need to choose  $\alpha,\beta, \epsilon>0$, 
with $\al$ as small  as possible,  so that
\[
\frac{1}{\sqrt{2}} b(k)e^{-2(\beta + \epsilon)^2\al k}> a(k)(1+\sqrt{\al k})e^{-c k}, 
\]
for $k$ large enough, or equivalently, 
\begin{equation}\label{messqaz2}
  2(\beta + \epsilon)^2\al  < c.  
\end{equation}
Note that $b(k)$ decays polynomially
($b(k) = \Omega(\frac{1}{k^{3/2}})$) and $a(k)$ increases logarithmically ($a(k) = O(\log{k})$), hence the effect 
of $\frac{1}{\sqrt{2}} b(k)$ and $a(k)(1+\sqrt{\al k})$  
vanishes  for $k$ large enough if $2(\beta + \epsilon)^2\al < c$.   

We are free to choose $\alpha, \beta, \epsilon>0$ as long as $\beta+\epsilon < 1$ and conditions 
(\ref{messqaz0}) and (\ref{messqaz}) of Lemma \ref{eee5} are satisfied.
We can ignore (\ref{messqaz0})  since it is implied by  
the condition $\beta+\epsilon < 1$. We can also ignore  (\ref{messqaz}) since (\ref{messqaz2}) is 
stronger as 
(\ref{messqaz}) is equivalent to $c>0$.
Writing (\ref{messqaz2}) as   $\al > h(\beta, \epsilon)$, where 
\[
 h(\beta, \epsilon) = 
 \frac{\log{(2+\e)}-\log{\beta}}{2(1-(\beta+\epsilon)^2)},
\]
we see that can attain any value of $\al >  \al^*$,  where 
$\al^* = \min   h(\beta, \epsilon)$, 
over the choice of $\beta, \epsilon >0$ such that $\beta + \epsilon \leq 1$.
Numerical evaluation shows that $\al^* \approx 0.9232$.  
For instance, for $\beta = 0.5204$ and
$\epsilon = 0.004$, we have 
 $h(\beta, \epsilon) \approx 0.9299$.  Setting $\al = 0.93$, we get  (\ref{shmuck}), and hence  
(\ref{thmainconc}), i.e., 
 $\E_n(k ,\sqrt{\al kn}) = 0$, for $k$ and $n$ sufficiently large. 
\finito

\subsection{Proof of Lemma \ref{eee5}}\label{eee5p}
The proof uses the following estimates. 

\begin{theorem}     \label{demlt}  
  {\bf (de Moivre-Laplace normal approximation of the binomial; see \cite{Fel68}, p. 184)}  
    Let $\delta: \N \rightarrow \N$  be such that $\delta(n) = o(n^{2/3})$.  Then,
    for each $\e>0$, there exists $n_0>0$ such that for each $n\geq n_0$ and 
    each integer $w$ such that   $|w-n/2|\leq \delta(n)$, we have 
\begin{equation*} 
(1-\e)  \sqrt{\frac{2}{\pi n}}e^{-2\frac{(w - n/2)^2}{n}}   \leq   B_n(w) \leq  (1+\e) \sqrt{\frac{2}{\pi n}}e^{-2\frac{(w - n/2)^2}{n}}. 
\end{equation*}
\end{theorem}
  \begin{theorem}
    [Hoeffding's inequality \cite{Hoe63}] \label{HoIneq} 
    Let $X_1, \ldots, X_n$ be independent random variables such that
    $a_i \leq X_i \leq b_i$, for each $i$. Then,  for each $t\geq 0$,
    \[
           Pr[\sum_i X_i - E[X_i] \geq t] \leq e^{-2 \frac{t^2}{\sum_i (b_i - a_i)^2}}.
           \]
\end{theorem} 
  Therefore, using Theorem \ref{demlt}, assume that $n$ is sufficiently large so that
  the following bounds 
  hold    for all  $w$ such that $|w-n/2|\leq  \frac{n^{2/3}}{\log^{1/2}{n}}$: 
  \begin{eqnarray}
  B_n(w) &\geq &\frac{1}{\sqrt{2 n}}e^{-2\frac{(w - n/2)^2}{n}}\label{dll}\\
    B_n(w) &\leq & \frac{1}{\sqrt{n}}e^{-2\frac{(w - n/2)^2}{n}}\label{dlu}.
\end{eqnarray}
To upper bound $B_n(w)$ for  $w> n/2+ \frac{n^{2/3}}{\log^{1/2}{n}}$, we
use  the following weak consequence of   Hoeffding's inequality. It follows from
Theorem \ref{HoIneq} that  for each integer $w\in [0:n]$, 
         \begin{equation}\label{Hoeff}
           B_n(w)  \leq e^{-2\frac{(w - n/2)^2}{n}}. 
         \end{equation}
               
         We use the lower bound (\ref{dll})
         to estimate $B_n\left(\floor{ n/2+ (\beta+\epsilon)\sqrt{         \al    k n}}\right)$.
 We use the upper bound         (\ref{dlu}) 
 to estimate the summation in (b) 
 for $n/2+\sqrt{\al kn} < w \leq n/2+\frac{n^{2/3}}{\log^{1/2}{n}} $. For $w> n/2+ \frac{n^{2/3}}{\log^{1/2}{n}}$, we use (\ref{Hoeff}).

          \paragraph{
            Proof of (a)}
          We have
\[
\frac{2}{n}     \left(\floor{ n/2+ (\beta+\epsilon)\sqrt{    \al k n}}-\frac{n}{2}\right)^2 \leq 2(\beta + \epsilon)^2 \al k.  
   \]
   Moreover, since $k \leq \frac{n^{1/3}}{\log^2{n}}$,
   \[ 
   \left|\floor{ n/2+ (\beta+\epsilon)\sqrt{\al   k n}}-\frac{n}{2}\right| \leq (\beta+\epsilon) \sqrt{\al kn}
   \leq (\beta + \epsilon)\sqrt{\al }
   \frac{n^{2/3}}{\log{n}}
      \leq 
     \frac{n^{2/3}}{\log^{1/2}{n}},
     \]
   for $n$ large enough.  
   It follows that
   \[   
  B_n\left(\floor{ n/2+ (\beta+\epsilon)\sqrt{\al k n}}\right) \geq  \frac{1}{\sqrt{2 n}}e^{-2(\beta + \epsilon)^2\al k }.
   \]
   
\paragraph{
  \bf Proof of (b)}   
          Decompose 
\[
\sum_{n/2+\sqrt{\al kn}< w  \leq n} \left(\frac{2(w-n/2)+\e\sqrt{\al kn}}{\beta  \sqrt{\al k n}}
\right)^{k} B_n(w ) = A + B,
\]
where 
\begin{eqnarray*}
  A &\defeq& \sum_{n/2+\sqrt{\al kn}< w  \leq n/2+\frac{n^{2/3}}{\log^{1/2}{n}}} \left(\frac{2(w-n/2)+\e\sqrt{\al kn}}{\beta\sqrt{\al k n}}  \right)^{k}  B_n(w ) \\
B &\defeq& \sum_{n/2+\frac{n^{2/3}}{\log^{1/2}{n}}<w\leq n} \left(\frac{2(w-n/2)+\e\sqrt{\al kn}}{\beta\sqrt{ l n}}\right)^{k} B_n(w ). 
\end{eqnarray*} 
We will argue that $A\leq \sqrt{\al k}  e^{-ck}$ and $B \leq e^{-ck}$, for sufficiently large $n$ and $k$.

If follows from  (\ref{dlu}) and (\ref{Hoeff}) that 
\begin{eqnarray*}
A &\leq& \frac{1}{\sqrt{ n}}
\sum_{w>n/2+\sqrt{\al kn}} \left(\frac{2(w-n/2)+\e\sqrt{\al kn}}{\beta\sqrt{ l n}}\right)^{k}  e^{-2\frac{(w - n/2)^2}{n}}\\
B &\leq&  
\sum_{w> n/2+\frac{n^{2/3}}{\log^{1/2}{n}}} \left(\frac{2(w-n/2)+\e\sqrt{\al kn}}{\beta\sqrt{ l n}}\right)^{k}  e^{-2\frac{(w - n/2)^2}{n}}.
\end{eqnarray*} 
Let $x = \frac{w-n/2}{\sqrt{\al k n}}$ and note that $x > 1 $ for $w>n/2+\sqrt{\al kn}$. 
Thus 
\begin{eqnarray*}
 \left(\frac{2(w-n/2)+\e\sqrt{\al kn}}{\beta\sqrt{\al k n}}\right)^{k}  e^{-2\frac{(w - n/2)^2}{n}}  
 &=&   e^{-2 \al k\left(x^2-\frac{1 }{2\al }\log{\frac{2x+\e}{\beta}}\right)}. 
\end{eqnarray*}
For all $x\geq  1$, we have 
$\log{ \frac{2x+\e}{\beta}}\leq \log{\left(\frac{2+\e}{{\beta}}     x\right)}\leq
 \left(\log{\frac{2+\e}{\beta}}\right)x^2$, 
 where  the last inequality holds for all $x \geq 1$ 
 iff $\frac{2+\e}{{\beta}}\geq \sqrt{e}$~~\footnote{Let $a = \frac{2+\e}{{\beta}}$. The  slopes at $1$  of $\log(a x )$ and $(\log{a}) x^2$  
 are $1$ and $2\log{a}$, respectively.
 Thus, to guarantee that  $\log(a x )\leq (\log{a}) x^2$ for all $x\geq 1$,   we need $1\leq  2 \log{a}$, i.e., $a \geq \sqrt{e}$.}, which is guaranteed by condition 
(\ref{messqaz0}).   
It follows that,  for all $x > 1$, 
\[
2 \al k\left(x^2-\frac{1}{2\al}\log{\frac{2x+\e}{{\beta}}}\right)
\geq 2 \left(\al - \frac{1}{2}\log{\frac{2+\e}{{\beta}}}
  \right) k  x^2
= c \frac{(w-n/2)^2}{\al n}, 
\]   
where $c  =  2\left(\al - \frac{1}{2}\log\frac{2+\e}{{\beta}}\right)$.
Note that condition (\ref{messqaz}) says that $c>0$.
Therefore,
\begin{eqnarray*}
A &\leq& \frac{1}{\sqrt{n}}
\sum_{w>n/2+\sqrt{\al kn}} e^{-c\frac{(w - n/2)^2}{\al n}}\\
B &\leq&  
\sum_{w> n/2+
\frac{n^{2/3}}{\log^{1/2}{n}}
} e^{-c\frac{(w - n/2)^2}{\al n}}.
\end{eqnarray*}
Now, in general, for any $y\geq 1$ and any $a>0$, we have 
\begin{eqnarray*}
\sum_{w>n/2+y} e^{-\frac{(w-n/2)^2}{a}}
&\leq&  \int_{n/2+y-1}^\infty e^{-\frac{(u-n/2)^2}{a}} du =\sqrt{a}\int_{\frac{y-1}{\sqrt{a}}}^\infty e^{-z^2} dz\\
&\leq& (y-1)\int_{\frac{y-1}{\sqrt{a}}}^\infty z e^{-z^2} dz =  \frac{y-1}{2} e^{-\frac{(y-1)^2}{a}} \leq  \frac{y}{2}
e^{-\frac{(y-1)^2}{a}}.
\end{eqnarray*}
It follows that
\[
A \leq \frac{\sqrt{\al k n}}{2\sqrt{n}}  e^{- \frac{c(\sqrt{\al kn} - 1)^2}{\al n}}
=  
\frac{1}{2} e^{\frac{c(2\sqrt{\al k n}-1)}{\al n}}
\sqrt{\al k}
e^{-ck}
\leq \sqrt{\al k} e^{-ck},  
\]
for $n$ large enough. 
The  last inequality holds for $n$ large enough since
$\frac{\sqrt{kn}}{n} \leq \frac{1}{n^{1/3}\log{n}}$ as 
$k \leq \frac{n^{1/3}}{\log^2{n}}$.   
Finally, 
\[
B \leq \frac{n^{2/3}}{2\log^{1/2}{n}}
e^{- c\frac{\left(\frac{n^{2/3}}{\log^{1/2}{n}} - 1\right)^2}{\al n}}
\leq
n^{2/3} 
e^{- \frac{c}{2\alpha}\frac{n^{1/3}}{\log{n}}} 
\leq n^{2/3} 
    e^{-  \frac{c}{2\alpha }k \log{n}} \leq e^{- ck }, 
\]  
for  $n$ and $k$ large enough, where  the inequality before the last holds because $k \leq \frac{n^{1/3}}{\log^2{n}}$.

\section{Conclusion}\label{conc}

We conclude with the following  open questions:

\begin{itemize}
  \item
As mentioned in the introduction, Tiet\"{a}v\"{a}inen's bound is not tight for linear codes with sufficiently large dual distance $d$ in the $d = \Theta(n)$ regime.
Is this also the case for $k$-wise independent distributions?  Note that 
in the aforementioned papers   \cite{SS93} - \cite{AB02},  the  techniques which improve on Tiet\"{a}v\"{a}inen's bound in the $d = \Theta(n)$ regime 
are  specific to linear codes and do not  seem applicable to $k$-wise independent distributions. 

\item
Is the upper bound  $\frac{n^{1/3}}{\log^2{n}}$ on $k$ in Theorem \ref{mainth} an artifact of the proof? i.e.,  does the statement hold  for larger values of $k$? 
The answer is not clear since if we ignore the log term, 
the requirement of $k = o(n^{1/3})$  has two independent origins
in the proof. The first is conditions   (\ref{qaz1}) and (\ref{qaz2}) in the analysis of quantized Chebyshev sequences in the proof of Lemma \ref{implem}. The second is  de Moivre-Laplace normal approximation of the binomial   in the proof of Lemma \ref{eee5}.

\item
Theorem \ref{mainth} implies the existence a $k$-wise independent probability distribution on $\{0,1\}^n$ whose covering radius is at least  $\frac{n}{2}-\sqrt{kn}$ if
$k \leq \frac{n^{1/3}}{\log^2{n}}$ and $k$ and $n$ are large enough.   
Can such distributions be supported by linear codes or are they 
intrinsically non-linear?  
That is,
assuming that $d = w(1)$ and $d=o(n)$, is there 
an $\F_2$-linear block-length-$n$ code  with  dual distance $d$ and  covering radius at least $\frac{n}{2}-\Theta(\sqrt{dn})$? 
\end{itemize}

\end{document}